\documentclass[pra,showpacs,preprint,nobibnotes,nofootinbib,floatfix]{revtex4}

\usepackage{amsmath}
\usepackage{amssymb}
\usepackage{bbold}
\usepackage{graphicx}
\usepackage{amsthm}

\DeclareMathOperator{\Tr}{\rm Tr}

\theoremstyle{plain}
\newtheorem{theorem}{{Theorem}}

\begin{document}

\title{An Entropy Inequality}

\date{\today}

\author{Meik Hellmund}
\affiliation{Mathematisches Institut, Universit{\"a}t Leipzig,
Johannisgasse 26, D-04103 Leipzig, Germany}
\email{Meik.Hellmund@math.uni-leipzig.de}

\author{Armin Uhlmann}
\affiliation{Institut f{\"u}r Theoretische Physik, Universit{\"a}t Leipzig,
Vor dem Hospitaltore 1, D-04103 Leipzig, Germany}
\email{Armin.Uhlmann@itp.uni-leipzig.de}

\begin{abstract}
Let $S(\rho)=-\Tr (\rho\log\rho)$ be the von Neumann entropy of an
$N$-dimensional quantum state $\rho$ and
 $e_2(\rho)$  the second elementary
symmetric polynomial of the eigenvalues of $\rho$. We prove the
inequality
\[
S(\rho) \;\le \;  c(N) \; \sqrt{e_2(\rho)}   \; 
\]
where $c(N)=\log(N) \, \sqrt{\frac{2N}{N-1}}$. This generalizes
an inequality given by Fuchs and Graaf \cite{fuchsgraaf} for the case of one
qubit, i.e., $N=2$. 
Equality is achieved if and only if $\rho$ is either a pure
or the maximally mixed state. This   inequality delivers new 
bounds for quantities of interest in quantum information theory, such as 
upper bounds  for  the minimum output entropy and the entanglement
of formation as well as a lower bound for the Holevo channel capacity.
\end{abstract}

\pacs{03.67.-a,   
03.67.Mn          
}
\maketitle

\section{Introduction}

\begin{figure}[Hht]
  \centering
 \includegraphics[scale=.3,angle=-90]{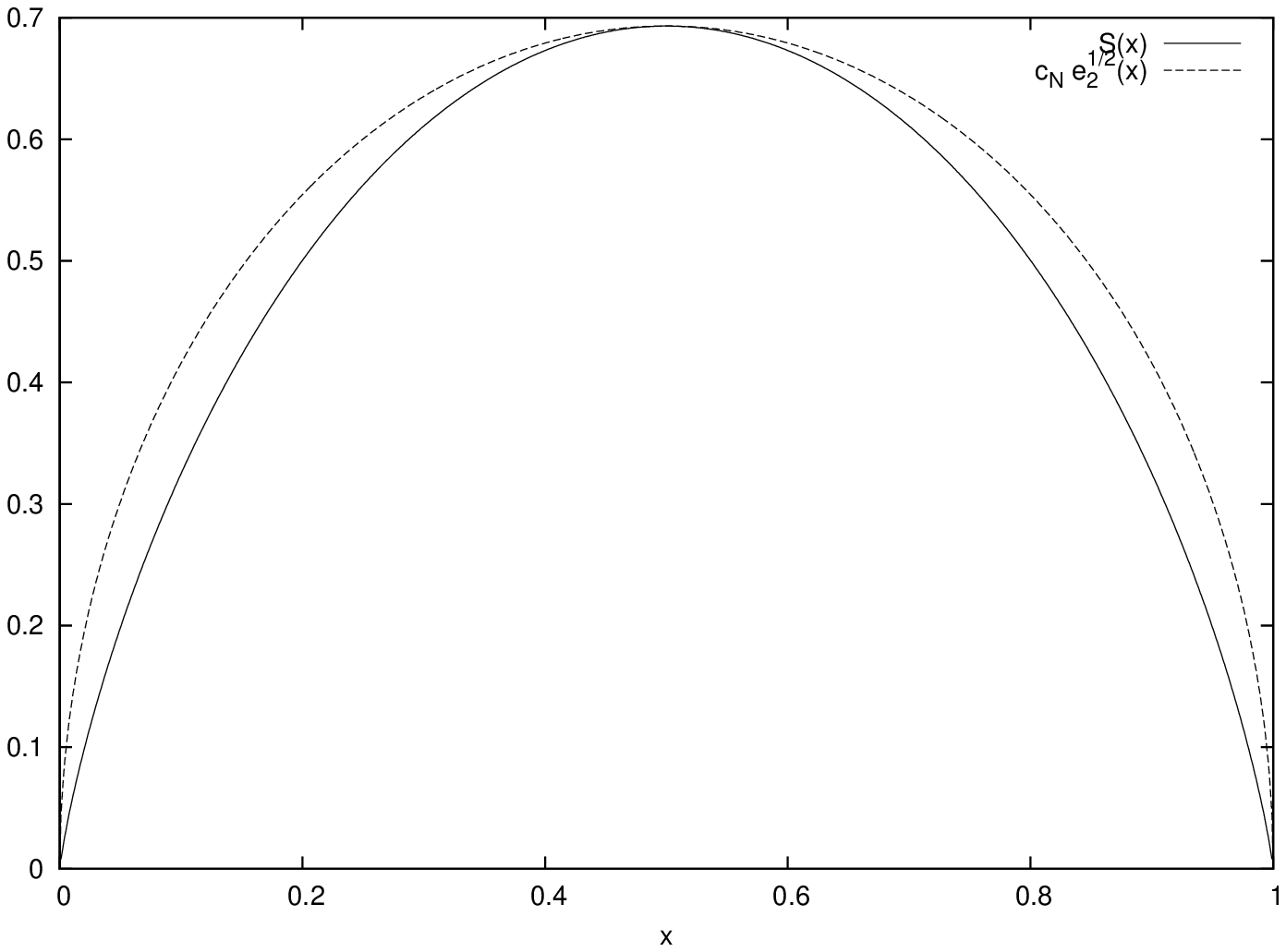}
  \caption{The Fuchs-Graaf inequality.}
  \label{fig:1}
\end{figure}

Let $\rho$ be a density matrix of a qubit with eigenvalues 
$x$ and $1-x$  ($0\le x\le 1$).
Its von Neumann entropy is given by
\begin{equation}
  \label{eq:1}
  S(\rho)  = \eta(x) + \eta(1-x)
\end{equation}
where the abbreviation $\eta(x):= -x \,\log x$
with  $\eta(0)=0$ is used.\footnote{Our formulas are valid for arbitrary
  bases of the logarithm. The natural logarithm is used in the figures.}
In \cite{fuchsgraaf} Fuchs and Graaf stated the inequality
\begin{equation}
  \label{eq:2}
  S(\rho) \le 2 (\log 2) \sqrt{x(1-x)}
\end{equation}
which  can be read off from figure \ref{fig:1}.

To gain the desired extension of eq.~(\ref{eq:2}) to $N$-dimensional
quantum systems, we  at first
observe that its right hand side is 1-homogeneous in $x_1 = x$,
$x_2 = 1-x$.
Therefore
after replacing $S$ by $(\Tr \rho) S(\rho [\Tr \rho]^{-1} )$,
the inequality becomes valid for all positive matrices, not only for density
matrices satisfying $\Tr \rho =1 $. 

Accordingly, we define
for any positive hermitian $N\times N$-matrix
 $\rho$ with eigenvalues  $x_i$
 \begin{eqnarray}
   \label{eq:3}
   S_1(\rho) &=&
    (\Tr \rho)  S(\frac{\rho}{\Tr \rho})
          = \sum \eta(x_i) - \eta(\sum x_i)\\
          e_2(\rho) &=& \frac{1}{2}  \left((\Tr\rho)^2-\Tr\rho^2\right)=
\sum_{i<j} x_i x_j
 \end{eqnarray}
The homogenized entropy  $S_1$ is of degree one,
$S_1(\lambda \rho) = \lambda S_1(\rho)$, and it clearly coincides
with $S$ at density matrices. $S_1$ is non-negative, concave, and
super-additive on the cone of positive matrices,
see for instance \cite{Boyd04}.
Similarly, $\sqrt{e_2}$  is of degree one.  It is concave and
super-additive for positive matrices.
The two functions eq.~(\ref{eq:3}) are bounded from above according to
\begin{eqnarray}
  \label{eq:5}
  S(\rho) &\le& \frac{\Tr\rho}{N} \; S(\mathbb 1) =  \log(N) \,\Tr\rho\\
  e_2(\rho) &\le& \left(\frac{\Tr\rho}{N}\right)^2 \, e_2(\mathbb 1) =
 \frac{N-1}{2N} \, (\Tr\rho)^2
\end{eqnarray}
 The central result of our paper is as following
 \begin{theorem}
   For all positive semi-definite $N\times N$ matrices $\rho$ we have
   \begin{equation}
     \label{eq:6}
     S_1(\rho) \; \le\; c_N \, e^{1/2}_2(\rho)
 , \qquad\text{where}\quad
 c_N = (\log N) \sqrt{\frac{2N}{N-1}}
\; .
   \end{equation}
Equality is achieved if and only if either $\rho$ is of rank one
(and  both sides of the inequality vanish)
or if $\rho$ is proportional to $\mathbb 1$.
 \end{theorem}
As an illustration of the theorem we show the difference
between the right and the left
hand side of this inequality for the case $N=3$ and $\Tr\rho=1$.
The difference vanishes at the corners (pure states) and at the center
(maximally mixed state). It takes its maximum along the edges, i.e., for
rank 2 states.

\begin{figure}[Hht]
  \centering
\includegraphics[scale=.8]{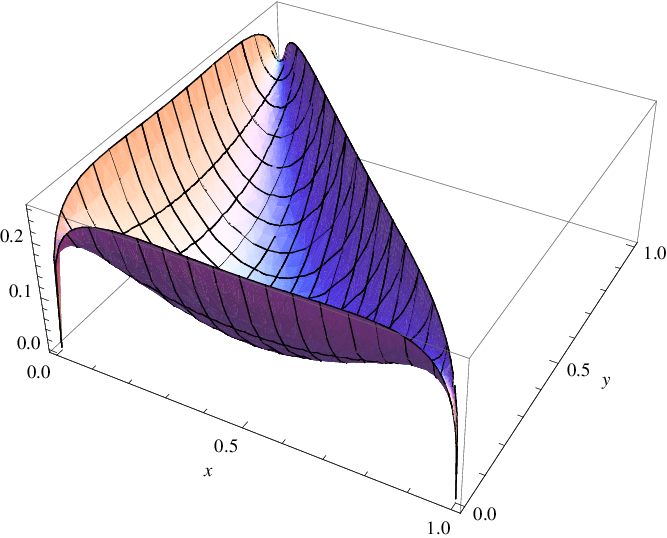}
  \caption{The difference $c_N e_2^{1/2}(\{x_i\})-S(\{x_i\})$ in the case $N=3$.
The eigenvalues are parameterized by $x_1=x, x_2=y, x_3=1-x-y$.  So, $\rho>0$
corresponds to the triangle $x\ge 0, y\ge 0, x+y\le 1$.
}
  \label{fig:2}
\end{figure}

Before giving the proof of this theorem we  add some observations.

\noindent {\em Remark 1:} \, $c_N$ is strictly increasing with $N$.\\
{\em Remark 2:} \,
If $\rho$ is of rank $k$ then the operator is supported by a
$k$-dimensional subspace. Using this sub-space, we see that
eq.~(\ref{eq:6}) remains valid after replacing $N$ by the rank of $\rho$.
By this argument we see that it suffices to prove eq.~(\ref{eq:6})
for matrices with maximal rank.\\
{\em Remark 3:} \, Below we shall use the notation
\begin{equation} \label{eq:6a}
f(\rho) = f(x_1, \dots, x_N) = \frac{S_1(\rho)}{\sqrt{e_2(\rho)}}
\end{equation}
where $x_1, \dots, x_N$ denote the eigenvalues of $\rho$.
We will prove that this function takes its global maximum 
at $\rho=\lambda {\mathbb 1}$.   
Numerical checks (up to $N=8$) support the more general\\
{\bf Conjecture:} {\em This function $f$  is concave on the set of density
operators.}\\
 {\em Remark 4:} In \cite{mitchison},  Mitchison and Jozsa 
considered the entropy as function of the elementary symmetric polynomials
$e_2(\rho),\dots, e_n(\rho)$ defined by, e.g., 
\begin{equation}
  \label{eq:15}
  \det(\lambda {\mathbb 1}-\rho) = \lambda^N-e_1(\rho)\lambda^{N-1} +
  e_2(\rho)\lambda^{N-2} -\cdots + (-1)^N e_N(\rho)
\end{equation}
 They showed that $\frac{\partial S}{\partial
  e_k} > 0$ for all $2\le k\le n$ (and therefore
$\frac{\partial S_1}{\partial
  e_k} > 0$ for all $1\le k\le n$.)  
In the light of this it seems natural to ask for the possibility of 
other  estimates of the entropy,  
for instance by using higher symmetric polynomials.

\section{Proof}
Abbreviating $x=\sum_{m=1}^N x_m$, $N \geq 2$,
we consider the function eq.~(\ref{eq:6a}),
\begin{equation}
  \label{eq:4}
  f_N(x_1,\dots,x_N) = \frac{S_1(x_1,\dots,x_N)}
{e_2^{1/2}(x_1,\dots,x_N)}=
\frac{\sum \eta(x_i) -\eta(x)}{(\sum_{i<k}x_i x_k)^{1/2}} .
\end{equation}
According to remark 2 we have to ask for extrema on $x_m > 0$.
This implies $x > x_m$ for all $m = 1, \dots, N$.
We use $\frac{\partial}{\partial x_m} e_2 = x - x_m,$ \ \
$ \frac{\partial}{\partial x_m} \eta(x) = -1-\log(x), $ \ \
$\frac{\partial}{\partial x_m} S_1 = \log\frac{x}{x_m}$ to get
\begin{equation}
  \label{eq:7}
  \frac{\partial f_N}{\partial x_m} = e_2^{-1/2}\left(\log\frac{x}{x_m}\right)
  -\frac{1}{2} (x-x_m) e_2^{-3/2} S_1.
\end{equation}
We look for extrema of $f_N$ under the condition $x = const.$ They must obey
\begin{equation}
  \label{eq:8}
  \frac{\partial f_N}{\partial x_m} =
\lambda \frac{\partial x}{\partial x_m} = \lambda,\qquad m=1,\dots, N
\end{equation}
Now $f_N$ is homogeneous of degree zero and $x$ of degree one.
Therefore,
\begin{equation}
  \label{eq:9}
  \sum_m x_m\, \frac{\partial f_N}{\partial x_m} =0,
\quad
\sum x_m \frac{\partial x}{\partial x_m} = x \; .
\end{equation}
Hence, eqs.~(\ref{eq:8}) can have solutions only for $\lambda=0$.
Now eq.~(\ref{eq:8}) reads
\begin{equation}
  \label{eq:10}
  e_2^{-1/2} (\log x -\log x_m) =\frac{1}{2} (x-x_m) e_2^{-3/2} S_1,
\qquad  m=1,\dots,N
\end{equation}
or,
\begin{equation}
  \label{eq:11}
  \frac{\log x-\log x_m}{x-x_m} =\frac{1}{2} e_2^{-1} S_1, \qquad m=1,\dots,N
\end{equation}
Now $x > x_m$ for all $m$ by assumption. One knows that
\begin{equation}
  \label{eq:12}
  x \mapsto \frac{\log y-\log x}{y-x}
\end{equation}
is strictly decreasing for $y>x>0$. Therefore, all $x_m$ must be equal and
\begin{equation}
  \label{eq:13}
  x_m = \frac{x}{N},\qquad m=1,\dots,N
\end{equation}
It is easy to check that this extremum is a maximum and therefore
\begin{equation}
  \label{eq:14}
  f_N(x_1,\dots,x_N) \le \frac{S_1(\mathbb 1)}{e_2^{1/2}(\mathbb 1)} =
\sqrt{\frac{2N}{N-1}}\;\log N 
\end{equation}
As this maximum is increasing with $N$, we are done.

\section{Applications} 
Let $\Phi: \rho_{in} \mapsto \rho_{out}=\Phi(\rho_{in})$ 
be a channel or, more general, a trace preserving
positive map between two finite-dimensional quantum state spaces. 

\subsection{The minimum output entropy}
The minimum output entropy, $S_{\rm min}(\Phi)$, is
the minimum of $S(\Phi(\rho))$ where $\rho$ is running through
all density operators. Obviously, $S_{\rm min}(\Phi)$
is smaller than
the minimal value of $c_N \sqrt{e_2(\Phi(\rho))}$, where $\rho$
is any density operator. Because $\sqrt{e_2}$ is concave, its
minimum is attained on rank one projection operators, i.e., pure states.
By our theorem we get the estimate
\begin{equation} \label{eq: m1}
S_{\rm min}(\Phi) \leq (\log n) \sqrt{\frac{n}{n-1}}\;
\min_{|\psi\rangle}
\sqrt{1 - \Tr \Phi(|\psi \rangle\langle \psi|)^2} \; 
\end{equation}
where the minimum runs through all unit vectors $|\psi\rangle$
and $n$ is the maximal rank attained by density operators
of the form $\Phi(|\psi \rangle\langle \psi|)$.

\subsection{A bound for the entanglement of formation by the concurrence}
The entanglement of formation of a bipartite pure state is defined as
the von Neumann entropy of one of the subsystems
\begin{equation}
  \label{eq:17}
  E_F(\psi) = S\left(\Tr_B(\pi_\psi)\right)\quad\text{ where } \pi_\psi=
|\psi\rangle\langle\psi|
\end{equation}
and extends to  mixed bipartite states by the convex roof
construction
\begin{equation}
  \label{eq:18}
  E_F(\rho) =   \min_{\sum p_i \pi_i=\rho}\; \sum\, p_i\, S(\Tr_B(\pi_i))
\end{equation}
where the minimum is taken over all convex decompositions of $\rho$ into a
mixture of pure states $\pi_i$, see Bennett et al \cite{BenFucSmo96}.
Another important entanglement measure is the concurrence, originally 
introduced for 2-qubit systems, see  \cite{Woo01}
 for a review. A possible generalization to larger systems proposed by 
Rungta et al \cite{RBCHMW} makes again use of the second symmetric polynomial:
\begin{equation}
  \label{eq:18a}
  C(\rho) \;= \; 2 \min_{\sum p_i \pi_i=\rho}\;  \sum\, p_i\,
e_2(\Tr_B(\pi_i))^{1/2}
\end{equation}
By theorem 1 every right hand side sum of eq.~(\ref{eq:18}) can be
bounded by a multiple of that of eq.~(\ref{eq:18a}). This simple
argument provides
\begin{equation} \label{eq:18c}
E_F(\rho) \;\leq\;  (\log n)\sqrt{ \frac{n}{2(n-1)}}\;\; C(\rho)
\end{equation}
with $n = \max \, {\rm rank}[\Tr_B(\pi)]$ the maximal rank
attained by the partially traced out pure density operators.

\subsection{A bound for the Holevo quantity $\chi^*$ }
For a channel map $\Phi$  one considers the Holevo quantity
\begin{equation}
  \label{eq:19}
  \chi^*_\Phi(\rho) = S(\Phi(\rho)) - \min_{\sum p_i \pi_i=\rho}\;
 \sum\, p_i\, S(\Phi(\pi_i))
\end{equation}
and the $\Phi$-concurrence
\begin{equation} \label{eq:19a}
C_{\Phi}(\rho) = 2 \min_{\sum p_i \pi_i=\rho}\;
 \sum\, p_i\, e_2(\Phi(\pi_i))^{1/2}
\end{equation}
Completely similar to the reasoning above we get the inequality
\begin{equation} \label{eq:19c}
  \chi^*_\Phi(\rho)\; \geq\; S(\Phi(\rho)) -  \log( n)\,\sqrt{ \frac{n}{2(n-1)}}
\;\; C_{\Phi}(\rho) \; .
\end{equation}
Here $n$ is again the maximal rank of the matrices $\Phi(\pi)$ with pure
$\pi$.


\end{document}